\documentclass[superscriptaddress,aps,prb,preprint]{revtex4-2}
\usepackage{graphicx}
\usepackage{epstopdf}
\usepackage{tabularx}
\usepackage{bm,amsmath,amssymb}
\usepackage{color, soul}
\usepackage{dcolumn} 
\usepackage{multirow}

\newcommand{\etal}{{\em et al.\ }}
\newcommand{\ie}{\textit{i}.\textit{e}.}
\begin{document}

\title{Accurate force field of two-dimensional ferroelectrics from deep learning }

\date{\today}
\author{Jing Wu}
\thanks{These two authors contributed equally}
\affiliation{Fudan University, Shanghai 200433, China}
\affiliation{School of Science, Westlake University, Hangzhou, Zhejiang 310024, China}
\author{Liyi Bai}
\thanks{These two authors contributed equally}
\affiliation{School of Science, Westlake University, Hangzhou, Zhejiang 310024, China}
\affiliation{Institute of Natural Sciences, Westlake Institute for Advanced Study,Hangzhou, Zhejiang 310024, China}
\author{Jiawei Huang}
\affiliation{Zhejiang University, Hangzhou Zhejiang 310058, P.R. China}
\affiliation{School of Science, Westlake University, Hangzhou, Zhejiang 310024, China}
\author{Liyang Ma}
\affiliation{Fudan University, Shanghai 200433, China}
\affiliation{School of Science, Westlake University, Hangzhou, Zhejiang 310024, China}
\author{Jian Liu}
\affiliation{Optics and Thermal Radiation Research Center, Shandong University, Qingdao, Shandong 266237, China}
\affiliation{School of Energy and Power Engineering, Shandong University, Jinan, Shandong 250061, China}
\author{Shi Liu}
\email{liushi@westlake.edu.cn}
\affiliation{School of Science, Westlake University, Hangzhou, Zhejiang 310024, China}
\affiliation{Institute of Natural Sciences, Westlake Institute for Advanced Study,Hangzhou, Zhejiang 310024, China}
\affiliation{Key Laboratory for Quantum Materials of Zhejiang Province, Hangzhou Zhejiang 310024, China}
\begin{abstract}
The discovery of two-dimensional (2D) ferroelectrics with switchable out-of-plane polarization such as monolayer $\alpha$-In$_2$Se$_3$ offers a new avenue for ultrathin high-density ferroelectric-based nanoelectronics such as ferroelectric field effect transistors and memristors. The functionality of ferroelectrics depends critically on the dynamics of polarization switching in response to an external electric/stress field. Unlike the switching dynamics in bulk ferroelectrics that have been extensively studied, the mechanisms and dynamics of polarization switching in 2D remain largely unexplored. Molecular dynamics (MD) using classical force fields is a reliable and efficient method for large-scale simulations of dynamical processes with atomic resolution. Here we developed a deep neural network-based force field of monolayer In$_2$Se$_3$ using a concurrent learning procedure that efficiently updates the first-principles-based training database. The model potential has accuracy comparable with density functional theory (DFT), capable of predicting a range of thermodynamic properties of In$_2$Se$_3$ polymorphs and lattice dynamics of ferroelectric In$_2$Se$_3$. Pertinent to the switching dynamics, the model potential also reproduces the DFT kinetic pathways of polarization reversal and 180$^\circ$ domain wall motions. Moreover, isobaric-isothermal ensemble MD simulations predict a temperature-driven $\alpha \rightarrow \beta$ phase transition at the single-layer limit, as revealed by both local atomic displacement and Steinhardt's bond orientational order parameter $Q_4$. Our work paves the way for further research on the dynamics of ferroelectric $\alpha$-In$_2$Se$_3$ and related systems.        
\end{abstract}

\maketitle
\newpage
\section{Introductions}
Ferroelectric materials with electrically addressable polarization have enabled various technological applications such as non-volatile memory, ferroelectric field effect transistors, sensors, and actuators~\cite{Garcia09p81,Lu12p59,Wan19p1808606,Liao09p700,Si17p24,Wang06p2768}. The continuing demand for miniaturized electronics has inspired the seek of new ferroelectrics that exhibit reversible polarization at reduced dimensions. Thin films of ferroelectric perovskite such as PbTiO$_3$, BaTiO$_3$, and BiFeO$_3$ have been the focus of numerous investigations, revealing a plethora of intriguing emergent phenomena  that may hold the promise for next-generation electronics~\cite{kinyanjui14p221909}.  However, the depolarization effect in thin films often becomes detrimental when the film thickness is approaching the sub-100~nm regime~\cite{Pinnow2004pK13}. The functional properties of ferroelectric thin films also depend sensitively on the lattice mismatch between the film and the underlying substrate as well as the growth quality of ferroelectric/metal interface~\cite{Yin13p397,Martin16p16087}, causing scaling issue on an industry scale. The poor compatibility of perovskite ferroelectrics with silicon is another challenge hindering the development of ferroelectric memory technology~\cite{Pinnow2004pK13}. 

Two-dimensional (2D) van der Waals (vdW) ferroelectrics with 
ferroelectric order in atom-thick layers 
are candidate materials addressing above issues because of their superior properties, \ie, uniform atomic thickness, the absence of surface dangling bonds, and the easy integration with other vdW layered materials.
An increasing number of 2D ferroelectrics have been
discovered by experiments or by first-principles calculations based on density functional theory (DFT) in the past decade~\cite{Wu21p9229}. Wu \etal proposed to functionalize graphene-based materials with hydroxyl groups to break the inversion symmetry in 2D~\cite{Wu13p081406}. Chang \etal reported robust in-plane ferroelectriciy in SnTe thin films with thickness down to one unit cell~\cite{Chang16p274}. The phosphorene and phosphorene analogues such as GeS and GeSe monolayers were predicted to be multiferroics with coupled ferroelasticity and ferroelectricity at room temperatures~\cite{Wu16p3236}. 
In practice, 2D materials with out-of-plane polarization are desirable for high-density integration because of the easiness of lateral downscaling. 
Vertical ferroelectricity has been discovered in monolayer
CuInP$_2$S$_6$~\cite{Belianinov15p3808}, $\alpha$-In$_2$Se$_3$~\cite{Ding17p14956}, distorted 1$T$-MoS$_2$~\cite{Shirodkar14p157601,Bruyer16p195402}, P$_2$O$_3$~\cite{Luo16p8575}, and MoTe$_2$~\cite{Yuan19p1775}. For vdW bilayers, the interlayer sliding is recently recognized as a convenient route to out-of-plane polarization, applicable even to non-ferroelectric 2D materials such as $h$-BN and AlN~\cite{Li17p6382, Zhang21p165420,Yasuda21p6549}.

Recently, In$_2$Se$_3$ has attracted significant attention for its diverse structural polymorphs and physical properties. Bulk In$_2$Se$_3$ can crystallize in several phases such as $\alpha$(2H), $\alpha$(3R), $\beta$(3R), $\gamma$, $\delta$, $\kappa$, where $\alpha$ and $\beta$ are layered vdW structures with 2H and 2R denoting different stacking sequences, and the other phases exhibit three-dimensional bonding networks~\cite{Han14p2747}. The commonly accepted phase transformation of bulk In$_2$Se$_3$ is $\alpha \rightarrow \beta \rightarrow \gamma \rightarrow \delta$ with increasing temperature, though the exact phase transition temperature remains controversial~\cite{Popovi79p416, Osamura66p1848,Lutz88p83,Kupers18p11775}. Additionally, in-situ  transmission electron microscopy characterizations suggested the formation of metastable phases, $\alpha'$, $\beta'$, and $\gamma'$, during heating or cooling of $\alpha$, $\beta$, and $\gamma$~\cite{Han14p2747}, respectively, highlighting the complexity of this indium selenides. 

\begin{figure}  
\centering
\includegraphics[width=1.0\textwidth]{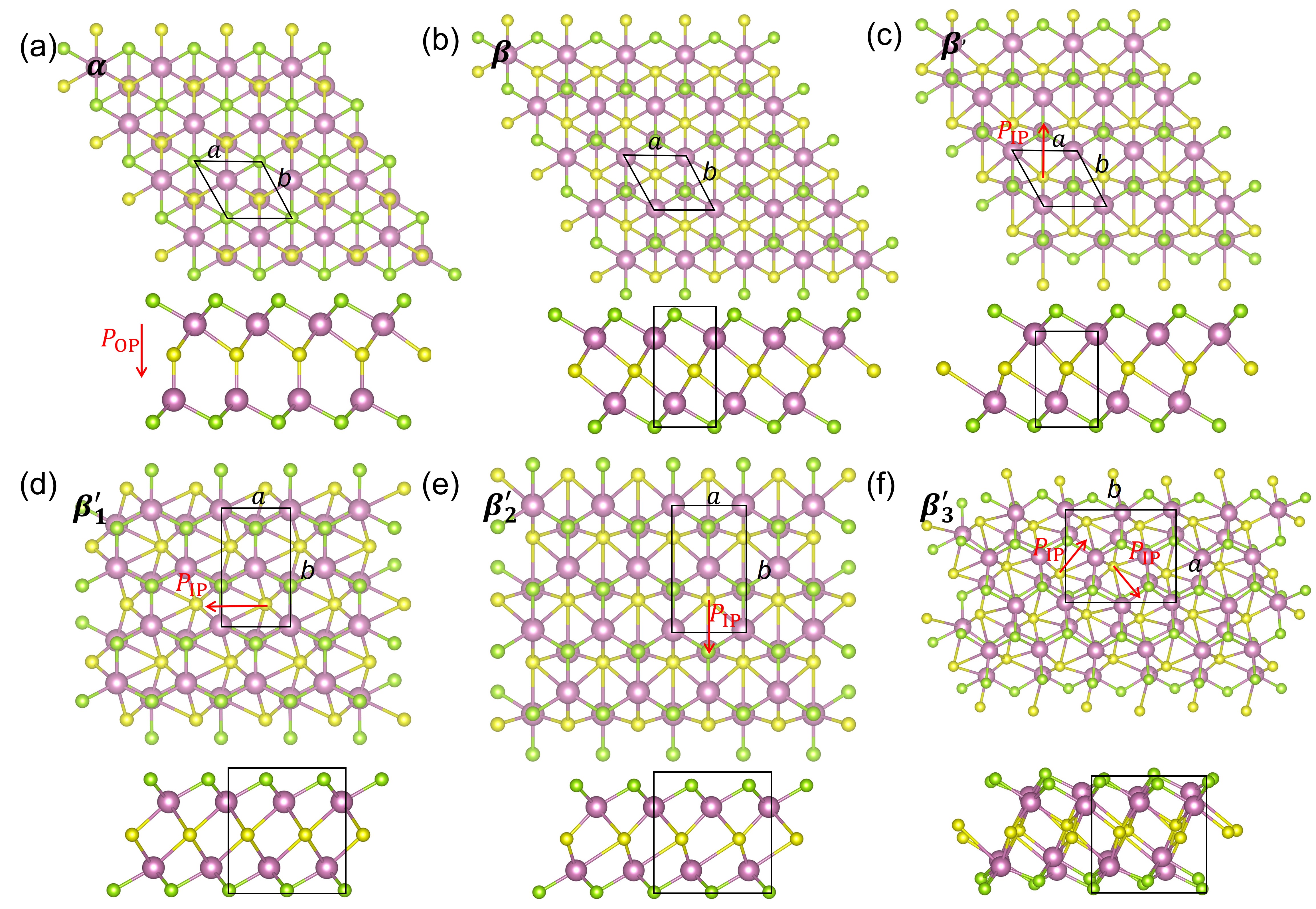} 
\caption{Top and side views of (a) $\alpha$,  (b) $\beta$,  (c) $\beta^{\prime}$, (d) $\beta_1^{\prime}$, (e) $\beta_2^{\prime}$, and (f) $\beta_3^{\prime}$ phases of monolayer In$_2$Se$_3$. The Se atoms in the middle sublayer are colored in yellow while the Se atoms in the top and bottom sublayers are colored in green. The red arrow represents the direction of local polarization. }
\label{fir:in2se3 phase}
\end{figure}

A typical monolayer in the layered phases of In$_2$Se$_3$ ($\alpha$ and $\beta$) is a covalently bonded quintuple layer (QL) consisted of five atomic planes in the sequence of Se-In-Se-In-Se. Atomically thin In$_2$Se$_3$ can adopt a few phases with distinct lattice distortions, as illustrated in Figure \ref{fir:in2se3 phase}. 
Ding \etal systematically investigated all possible atomic configurations within a QL with DFT and found that the $\alpha$-In$_2$Se$_3$ monolayer has the lowest energy and exhibits out-of-plane polarization ($P_{\rm OP}$) and in-plane asymmetry. The presence of $P_{\rm OP}$ in multilayered $\alpha$-In$_2$Se$_3$ nanoflakes have been confirmed experimentally~\cite{Cui18p1253,Xue18p4976,Xiao18p227601,Poh18p6340,Xue18p1803738}. Notably, through second harmonic generation spectroscopy and piezoresponse force microscopy, Xiao \etal reported a robust out-of-plane ferroelectricity in $\alpha$-In$_2$Se$_3$ crystals with thickness down to 3~nm and a high phase transition temperature ($T_C$) of 700~K for $\alpha \rightarrow \beta$ structural transition~\cite{Xiao18p227601}; Xue \etal reported ferroelectric hysteresis loops at the monolayer limit~\cite{Xue18p4976}. Structurally, the $\beta'$ phase with in-plane polarization ($P_{\rm IP}$) can be obtained by slightly displacing in-plane the central Se atoms of the nonpolar $\beta$ phase. Zheng \etal found that $\beta'$-In$_2$Se$_3$ has stable in-plane ferroelectricity up to 473~K in both bulk crystals and thin layers down to 45 nm~\cite{Zheng18peaar7720}. Zhang \etal discovered another phase at 77~K (named as $\beta'$ in the original paper but $\beta_3'$ in this work ) in In$_2$Se$_3$ with a thickness from more than 10 layers down to the monolayer~\cite{Zhang19p8004}. We note that $\beta_3'$-In$_2$Se$_3$ has canted in-plane dipoles resulting in a net in-plane polarization, a feature that is rare in bulk ferroelectrics and worthy of further investigations. Two metastable structure, $
\beta_1'$ and $\beta_2'$, can be derived from the unstable phonon modes of $\beta$-In$_2$Se$_3$ at the center of the Brillouin zone followed by the break of in-plane threefold rotational symmetry. The in-plane lattice distortions in $\beta_2'$-In$_2$Se$_3$ resemble those in the polar nanostripes of $\beta'$-In$_2$Se$_3$ superstructures recently observed in experiments, where the periodic antiparallel ordering between neighboring nanostripes gives rise to 2D antiferroelectricity~\cite{Xu20p047601}. It is noted that  $\alpha$, $\beta$ and $\beta'$ phases have hexagonal symmetry whereas $\beta_1'$, $\beta_2'$, and $\beta_3'$ phases are in the orthorhombic crystal family. 

No different from their 3D counterparts, the functionality of 2D ferroelectrics depends critically on the response of the polarization to the external stimuli such as thermal, stress, and electric field. Apparently, a high $T_C$ is favored for everyday applications. Previous studies on various perovskite ferroelectrics showed that $T_C$ will decrease with reducing film thickness because of the stronger depolarization effect~\cite{Streiffer02p067601,Glinchuk02p356}. 
In comparison, $\alpha$-In$_2$Se$_3$ nanoflakes exhibit an unusually high $T_C$ of $\approx$700~K. The coercive field and the switching speed dictate the power consumption and writing speed of ferroelectric memory, respectively. The mechanisms and kinetics of ferroelectric switching in 3D ferroelectrics have been the focus of numerous experimental and theoretical studies. The commonly used models for domain switching are Kolmogorov-Avrami-Ishibashi (KAI) model~\cite{Kolmogoroff37p335,Avrami39p1103,Ishibashi71p506},  nucleation limited switching (NLS) model~\cite{Jo07p267602,Tagantsev02p214109}, and homogeneous Landau switching model~\cite{Hoffmann19p464}, the applicability of which depends largely on the sample dimensions. However, 
it remains questionable whether the domain switching models developed for 3D ferroelectrics are applicable at reduced dimensions. In particular, the domain wall in 2D ferroelectrics is essentially a 1D interface, which might have creep and depinning behavior fundamentally different from the 2D domain wall in bulk ferroelectrics~\cite{Pascal02p6241}. Studying the dynamics of ferroelectric order in 2D is therefore both fundamentally and technologically important. 

To observe and quantify the domain switching process in 2D ferroelectrics with sufficiently high spacial/time resolution remains a great experimental challenge. As a consensus, molecular dynamics (MD) is a suitable method to study the switching process at femtosecond and million atom scale~\cite{Liu16p360, Takenaka17p391}. However, a force field that accurately describes the interatomic interactions is the prerequisite of running a MD simulation. Such force field is often not available for newly discovered materials. Since the discovery of graphene in 2004, a large number of 2D materials have been discovered. Nevertheless, only a handful of force fields of 2D materials have been developed, \ie,  graphene~\cite{Stuart00p6472,Brenner02p783,Lindsay10p205441}, MoS$_2$~\cite{Liang09p245110}, and $h$-BN~\cite{Mortazavi15p13228,Rajan18p1584}, most of which are based on the reactive empirical bond-order (REBO) potential~\cite{Brenner02p783}. Such scarcity reflects the difficulty of developing classical force fields for 2D materials. For a many-body potential such as REBO that uses a highly sophisticated energy function with a large number of parameters, the fitting process is often tedious. Moreover, there is no guaranty that the energy function developed for a specific class of material ({\em e.g.}, hydrocarbon) is robust enough to represent the ``true" interatomic potential of a different class ({\em e.g.}, 2D ferroelectrics). Additionally, for a thermodynamic property, different force fields for the same material may predict drastically different values. For example, the thermal conductivity of graphene was calculated to be 870~\cite{Wei11p2653}, 709.2~\cite{Hong18p2637}, 350~\cite{Thomas10p045413}, and 3000 W mK$^{-1}$~\cite{Mortazavi15p1,Fan17p14} using force fields of Terosff~\cite{Tersoff89p5566}, AIREBO~\cite{Stuart00p6472}, REBO~\cite{Brenner02p783}, and optimized Tersoff~\cite{Lindsay10p205441}, respectively, presenting almost an order of magnitude variation.

Herein, using the Deep Potential Molecular Dynamics (DeePMD) method~\cite{Zhang20p107206}, we developed a deep neural network-based force field applicable to ferroelectric monolayer $\alpha$-In$_2$Se$_3$. The Deep Potential (DP) scheme represents the total energy as a sum of atomic energies, each parameterized with a deep neural network (DNN) taking atomic local environment as input (see details in ref.~\citenum{Zhang20p107206}). The resultant DP model reproduces the DFT results for a range of thermodynamic properties of various phases of monolayer In$_2$Se$_3$. The DP energy barriers for a few polarization switching processes agree reasonably well with DFT values. Particularly, the temperature-driven ferroelectric-paraelectric phase transition of monolayer $\alpha$-In$_2$Se$_3$ is captured by DPMD simulations, with a theoretical $T_C \approx650$~K matching with recent experimental value ($\approx 700$~K). Our work paves the way for further research on the dynamics of 2D ferroelectricity in monolayer $\alpha$-In$_2$Se$_3$ and related systems. We expect the current DP model can be systematically improved and extended, enabling large-scale simulations of interlayer sliding ferroelectricity and Moir\'e ferroelectricity of vdW multilayers. 

 \section{Computational Methods}
\subsection{Deep potential generator }
\begin{figure}  
\centering
\includegraphics[width=0.7\textwidth]{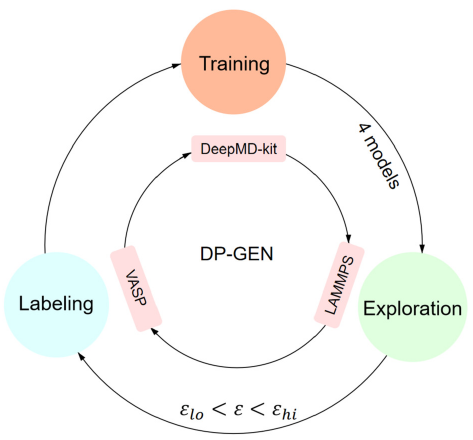} 
\caption{(a) DP-GEN workflow with training, exploration, and labeling steps forming a closed loop. At the training step, four models are trained using first-principles energies, atomic forces, and stress. The model deviation $\mathcal{E}$ is used as an error indicator for importance sampling, and only configurations satisfying $\mathcal{E}_{\rm lo} < \mathcal{E} < \mathcal{E}_{\rm hi}$ are labeled for DFT calculations.}
\label{fir:dpgen}
\end{figure}

In this work, we employ the deep potential generator (DP-GEN) scheme~\cite{Zhang19p023804} as implemented in \texttt{DP-GEN} package to develop the deep potential of In$_2$Se$_3$ with minimal human intervention. DP-GEN is a concurrent learning procedure with a closed loop consisted of three steps, exploration, labeling, and training, as shown in Figure \ref{fir:dpgen}. Additional details can be found in the original paper~\cite{Zhang19p023804} and our recent work on developing a DP model of HfO$_2$~\cite{Wu21p024108}. In the training step, four models are trained with different initial values of hyperparameters of DNNs  while using the same database of DFT energies and atomic forces. To gauge the necessity of DFT calculations for a configuration sampled from a DPMD run at the exploration step, the model deviation $\mathcal{E}$ is used, which is defined as the maximum standard deviation of the predictions of the atomic forces $\bm{F}_i$ from four different DP models,
\begin{equation}
\mathcal{E} = \max_i \sqrt {\left < || \bm{F}_i-  \left <  \bm{F}_i    \right > ||^2 \right > }
\end{equation}
where $i$ is the atomic index for atoms in a configuration and $\left< ...\right>$ denotes the average of four DP models. We introduce two thresholds, $\mathcal{E}_{\rm lo}$ and $\mathcal{E}_{\rm hi}$; only configurations satisfying $\mathcal{E}_{\rm lo} < \mathcal{E} < \mathcal{E}_{\rm hi}$ are labeled for DFT calculations. That is, the model deviation serves as an indicator for importance sampling.
For a configuration sampled from MD simulations at the exploration step, the four DP models in principle will exhibit nearly the same predictive accuracy ($\mathcal{E}<\mathcal{E}_{\rm lo}$) if the configuration is well represented by the current training database. In contrary, a configuration with $\mathcal{E}>\mathcal{E}_{\rm lo}$ indicates, heuristically, that it is ``outside" the configuration space of the training database, the DFT energy and atomic forces of which will be added to the database for training in the next cycle.
The introduction of $\mathcal{E}_{\rm hi}$ is to handle the exceptions due to highly distorted configurations (
resulting from low-quality DP models especially in the first few cycles of DP-GEN) or unconverged DFT calculations. When all sampled configurations have $\mathcal{E}< \mathcal{E}_{\rm lo}$, the ensemble of DP models is considered converged. Here we set $\mathcal{E}_{\rm lo}=0.12$~eV/\AA~and $\mathcal{E}_{\rm hi}=0.25$~eV/\AA.  

The smooth version of the DP model with an inner cutoff of 1~\AA~and an outer cutoff of 6~\AA~was adopted~\cite{Zhang18p4441} in this study, and the \texttt{DeePMD-kit} package~\cite{Wang18p178} was used for training. The embedding network has the ResNet-like architecture with a size of (25, 50, 100). The fitting network comprises three layers, each containing 240 nodes, with the loss function defined as 
\begin{equation}
L({p}_\epsilon, {p}_f, {p}_\xi) = {p}_\epsilon \Delta{\epsilon}^2 + \frac{p_f}{3N} \sum_i \left| \Delta{{\bm {F}}}_i \right|^2  + \frac{p_\xi}{9}  \left \| \Delta \xi \right \|^2
\end{equation}
where $\Delta$ denotes the difference between the DP results and the training data, $N$ is the number of atoms, $\epsilon$ is the energy per atom, ${\bm {F}}_i$ is the atomic force of atom $i$, and $\xi$ is the virial tensor divided by $N$. The parameters, $p_\epsilon$, $p_f$, and $p_\xi$, essentially tune the weights of energy, force, and stress information used in training, and can be changed even during the learning process~\cite{Wang18p178}. Here we increase both $p_\epsilon$ and $p_\xi$ from 0.02 at the beginning to 1 as the learning ends, while $p_f$ decreases from 1000 to 1. The number of batch and step of learning rate decay are set to 4000000 and 20000, respectively.

\subsection{Initial training database and exploration protocol}
The DP-GEN scheme allows for automatic, iterative, and efficient updates of the training database, making the force field development less impacted by the construction of the initial training database. Nevertheless, a carefully designed initial database with diverse configurations could speed up the reach of model convergence. The initial training database contains ground-state structures and configurations with random atomic/strain perturbations of bulk polymorphs (space group: $P\bar{3}m1$, $R3m$, and $P6_1$) and monolayer of different phases including $\alpha$, $\beta'$, $\beta_1'$, $\beta_2'$, and $\beta$ phases. We use a 20-atom slab model to represent the monolayer system and a 30-atom supercell to model the bulk. DFT calculations are carried out with the Vienna Ab initio Simulation (\texttt{VASP}) package~\cite{Kresse96p11169,Kresse96p15}, employing the projected augmented wave (PAW) method~\cite{Blochi94p17953,Kresse99p1758} and the generalized gradient approximation of Perdew-Burke-Ernzerhof (PBE)~\cite{Perdew96p3865} type for the exchange-correlation functional  as most of previous first-principles DFT studies of monolayer In$_2$Se$_3$ used PBE. A DP model trained with a PBE-based database allows us to directly compare DP predictions with previous DFT results. 
The main focus of this work is to develop an accurate model potential for monolayer In$_2$Se$_3$. For this reason, the vdW correction was not employed as it is likely not important for intralayer interactions dominated by covalent bonds. 

An energy cutoff of 700 eV and a $k$-point spacing of 0.3~\AA$^{-1}$ (corresponding to a $4\times4\times1$ $k$-point mesh for a 20-atom slab model) are sufficient to converge the energy and atomic force.  At the exploration step, the configuration space is sampled by running $NPT$ simulations at various temperatures (from 300 to 1200~K) at 0 and 1~kBar, respectively. 

\subsection{MD simulations of phase transition} 
The converged DP model of In$_2$Se$_3$ is used to investigate the temperature-driven phase transition of monolayer $\alpha$-In$_2$Se$_3$ by running $NPT$ MD simulations from 300~K to 1200~K with a temperature interval of 50~K. Similar to our previous study on HfO$_2$~\cite{Wu21p024108}, we found that the value of $T_C$ from MD exhibits supercell size effect. We thus used a sufficiently large 8000-atom supercell consisted of $40 \times40 \times 1$ unit cells to obtain the converged $T_C$. The MD trajectories are propagated using the velocity Verlet algorithm~\cite{Allen17book} and a time step of 1 fs, with temperature and pressure controlled by Nos\'e-Hoover thermostat and Parrinello-Rahman barostat, respectively, as implemented in \texttt{LAMMPS}~\cite{Plimpton95p1}. At each temperature, an equilibrium run of at least 10~ps is performed followed by a  60~ps production run that samples configurations for property analysis. At temperatures close to $T_C$, the production run is 500~ps to ensure enough sampling to obtain statistically reliable values of thermodynamics properties.

\section{Results}
\subsection{Fitting performance of DP model}
The final training database records 22600 monolayer configurations and 2163 bulk structures. The model deviation $\mathcal{E}$ indeed serves as an efficient indicator of importance sampling. The convergence was reached after 20 iterations during which more than 8 million configurations were sampled while only 7124 (0.09$\%$) configurations were deemed important for DFT calculations, substantially reducing the computational cost associated with DFT calculations.  Moreover, the automatic and iterative updates of the training database on the fly greatly alleviate the burden and reduce the bias of human interventions. 

\begin{figure}  
\centering
\includegraphics[width=1.0\textwidth]{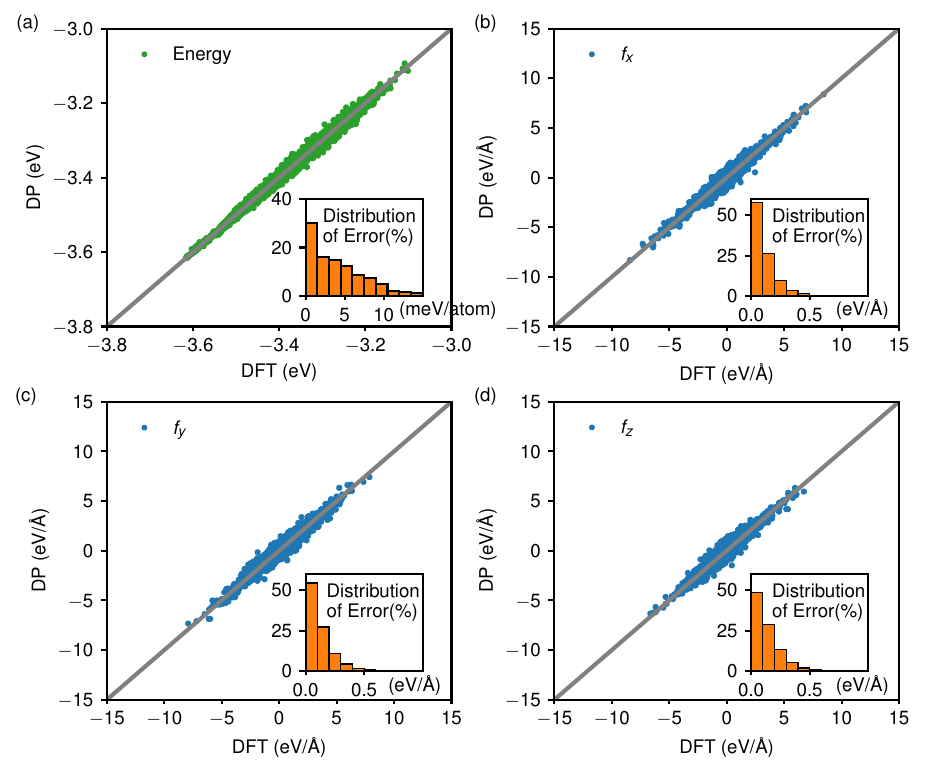} 
\caption{Comparison of DFT and DP (a) energy per atom and (b-d) atomic forces for configurations in the final training database. The insets provide the distribution of the absolute error.}
\label{fir:fitting}
\end{figure}

\begin{figure}  
\centering
\includegraphics[width=0.6\textwidth]{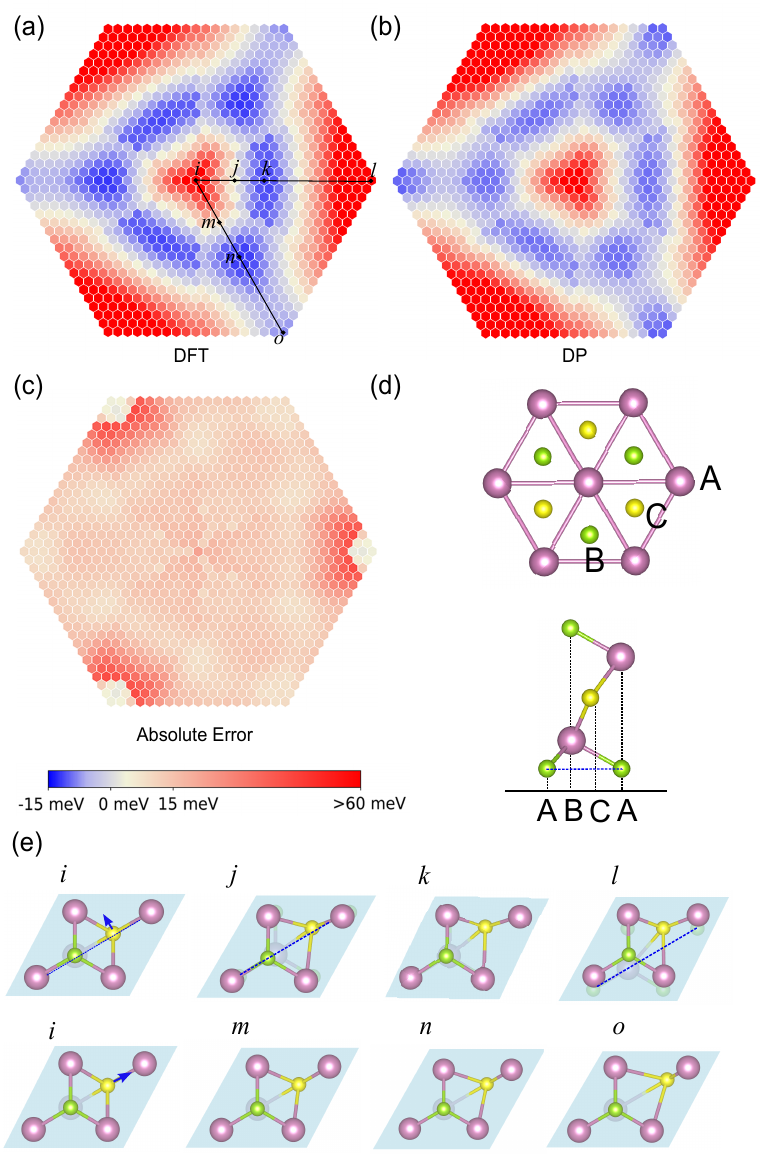} 
\caption{Two dimensional potential energy surface for the in-plane sliding of the central Se sublayer in the QL computed with (a) DFT and (b) DP. 
 As the structure has six-fold symmetry, the energy map was obtained by calculating the energies of configurations in the one sixth of the whole map area. (c) Absolute error in energy per 5-atom unit cell. (d) Top and side view of the QL used as the zero energy reference (denoted as fcc' in ref.~\citenum{Ding17p14956}). Atoms in each sublayer are arranged in one of the triangular lattices denoted as A, B, and C, respectively. (e) Configurations along two pathways highlighted in (a). For all the configurations, the central Se atom is fixed while all the other atoms are fully relaxed. 
}
\label{fir:map}
\end{figure}

Figure~\ref{fir:fitting} compares the energy per atom and atomic forces predicted by DFT and DP for all the structures, with the insets illustrating the absolute error distributions. The excellent representability of the DNN-based potential is evident from the small mean absolute error (MAE) of just 5.72 meV/atom for nearly 25000 configurations. Additionally, the training database contains a set of configurations that construct a 2D potential energy surface for the in-plane sliding of the central Se sublayer within the QL. The DFT and DP energy maps reported in Figure \ref{fir:map} show remarkable agreement with a MAE of 2.68~meV/atom. The capability to faithfully reproduce the energies and atomic forces of a diverse set of configurations gives the trained DP model a firm basis to accurately predict thermodynamic and kinetic properties of In$_2$Se$_3$.

\subsection{Predictions of static properties of In$_2$Se$_3$}
\begin{figure}  
\centering
\includegraphics[width=1.0\textwidth]{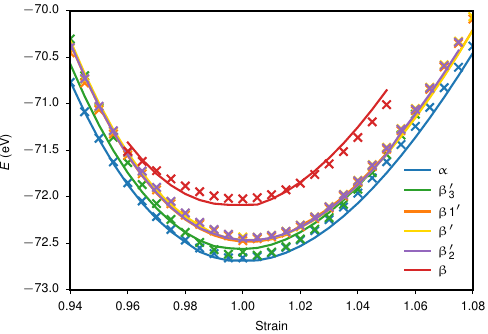} 
\caption{Energy-strain plots for monolayer In$_2$Se$_3$ of different phases subjected to equibiaxial deformations. Solid lines and cross points denote DFT and DP results, respectively.}
\label{fir:eos}
\end{figure}

Unlike a supervised learning task that deals with a fixed data distribution from which the validation dataset can be drawn to check generalization and overfitting problems, the accuracy of a DP model is better confirmed by comparing DP predictions with DFT and/or known experimental results not included in the training database~\cite{Wu21p024108}. 
We first check the predictive performance of the DP model by computing various thermodynamic properties. 
Table I compares the DFT and DP energies of different phases of monolayer In$_2$Se$_3$, demonstrating an excellent agreement. The DP model predicts the correct energetic ordering $E(\alpha) < E(\beta_{3}')< E(\beta_{1}') <E(\beta') <E(\beta_{2}') < E(\beta)$. In particular, the energy differences between those $\beta$-derived phases are within 2~meV/atom, all reproduced by the DP model. Figure \ref{fir:eos} presents the DP and DFT energy-strain plots for monolayers subjected to equibiaxial deformations. The DP model achieves a satisfactory accuracy in this test as well: the DFT and DP curves are nearly overlapping with each other over a wide strain range ($\pm6$\%). We note that the experimentally observed $\beta_3'$ phase, though not included in the training database, is described reasonably well by the DP model, hinting at some level of transferability of the DP model. 

\begin{figure}  
\centering
\includegraphics[width=1.0\textwidth]{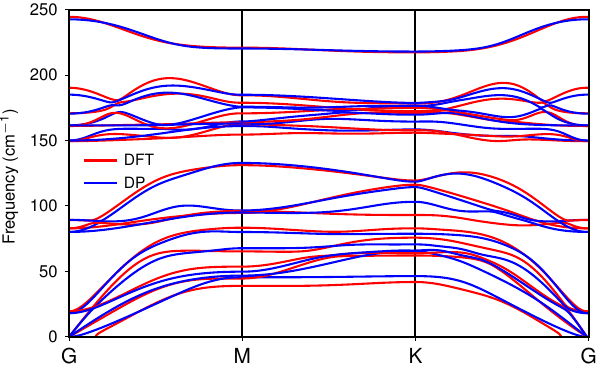} 
\caption{Phonon dispersion relations of monolayer $\alpha$-In$_2$Se$_3$ computed with DFT and DP.}
\label{fir:phonon}
\end{figure}

Accurate prediction of the phonon dispersion relationship is essential for quantitative evaluations of thermodynamic properties at finite temperatures. The lattice constant of $\alpha$-In$_2$Se$_3$ optimized with the DP model is $a=4.101$~\r{A}, within 1\% of the DFT value of 4.063~\r{A}. We then compute the phonon spectra of $\alpha$-In$_2$Se$_3$ with DFT and DP, respectively, using the frozen phonon approach as implemented in \texttt{Phonopy}~\cite{Togo15p1}. 
As shown in Figure \ref{fir:phonon}, the DP phonon spectrum of monolayer $\alpha$-In$_2$Se$_3$ show a decent agreement with DFT spectrum. It is often challenging to reproduce first-principles phonon dispersion relationships for a classical force field because of the demanding requirement to accurately fit the second-order derivatives of the potential energy surface. The confirmed accuracy of the DP model for predicting lattice dynamics highlights the robustness of the DP-GEN scheme that does not explicitly include perturbed structures generated within the frozen phonon approximation for phonon spectrum calculations. It is expected that the DP model can be further improved if all the perturbed supercells are included in the training database. 

\subsection{Domain switching}
\begin{figure}  
\centering
\includegraphics[width=0.8\textwidth]{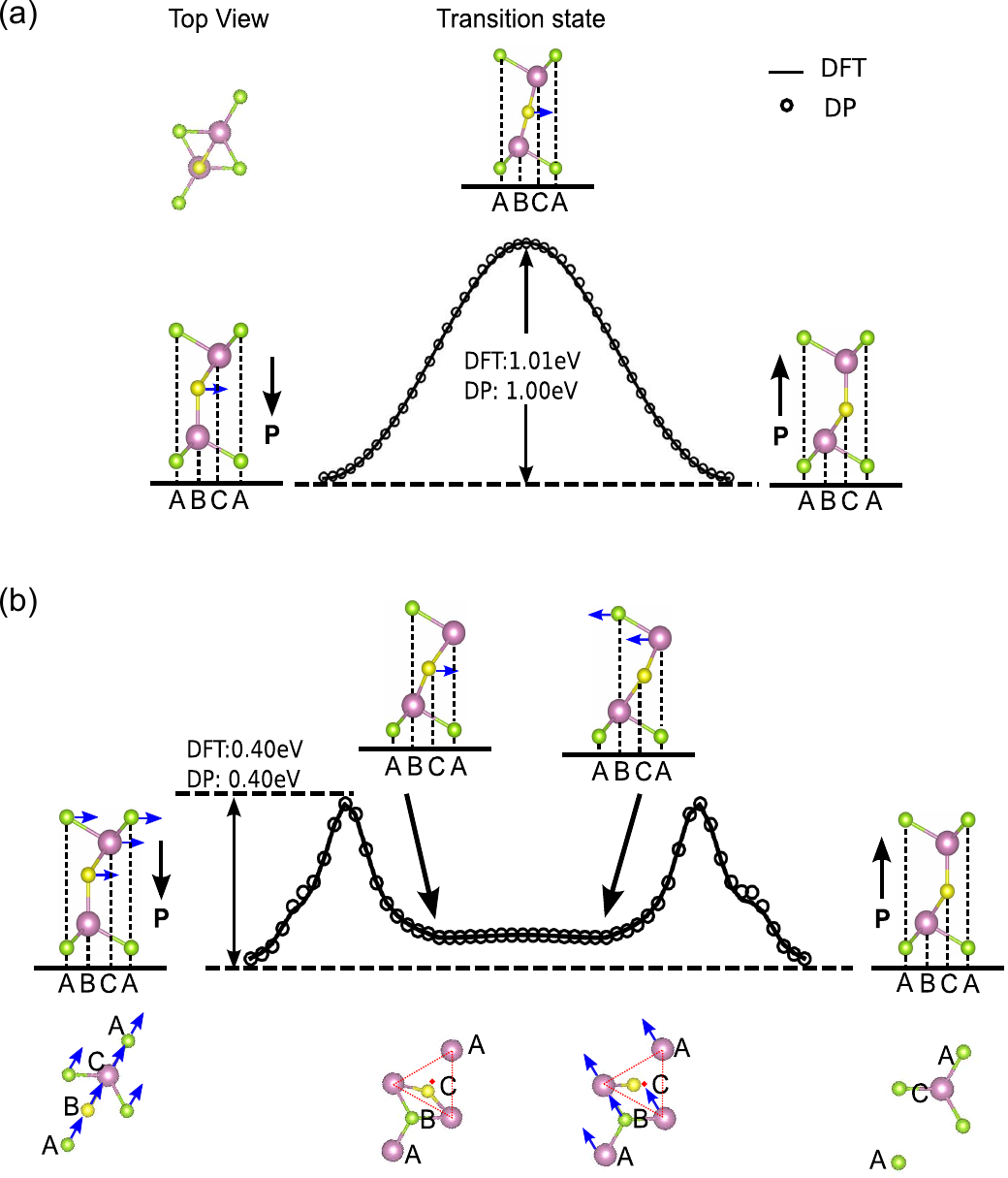} 
\caption{DFT and DP energy profiles for polarization reversal pathways in monolayer $\alpha$-In$_2$Se$_3$. (a) Direct switching of out-of-plane polarization via the displacement of the central Se sublayer. (b) Multistep switching via concerted motions of upper Se-In-Se layers. Only the top three sublayers, Se-In-Se, are shown in the top view. The flat region of the energy profile corresponds to the rotation of the middle Se sublayer around the ideal C point (red dot) by 60$^\circ$.}
\label{fir:path}
\end{figure}

In order to use MD to study polarization reversal at finite temperatures, it is necessary for the force field to accurately reproduce the DFT kinetic barriers of representative processes at 0~K.
Structurally, the out-of-plane polarization of $\alpha$-In$_2$Se$_3$ is due to the displacement of the central Se layer with respect to top and bottom In-Se layers. Ding \etal~identified two polarization reversal pathways with the nudged elastic band (NEB) method: the direct shifting of the central Se layer and a three-step concerted motion of upper Se-In-Se layers~\cite{Ding17p14956}, as shown in Figure~\ref{fir:path}. 
We determined the the minimum energy paths (MEPs) using the NEB technique implemented in the \texttt{USPEX}~\cite{Oganov06p244704,Lyakhov13p1172,Oganov11p227} code based on the switching trajectories proposed in ref.\citenum{Ding17p14956}. It is noted that the DFT calculations of energy and force required in the NEB method were performed using the same settings ({\em e.g.}, cutoff energy and $k$-point density) as those at the labeling step of the DP-GEN scheme.
The energies of configurations along the MEPs were then evaluated with the DP model.
As shown in  Figure \ref{fir:path}, the DFT and DP energy profiles of the two polarization reversal pathways are nearly identical with a small MAE of 1.68~meV/atom. In particular, the reproduction of the DFT energy profile for the complicated three-step switching process (Fig.~\ref{fir:path}(b)) that involves highly distorted intermediate structures manifests the quantitative nature of the DP model. 
  
\begin{figure}  
\centering
\includegraphics{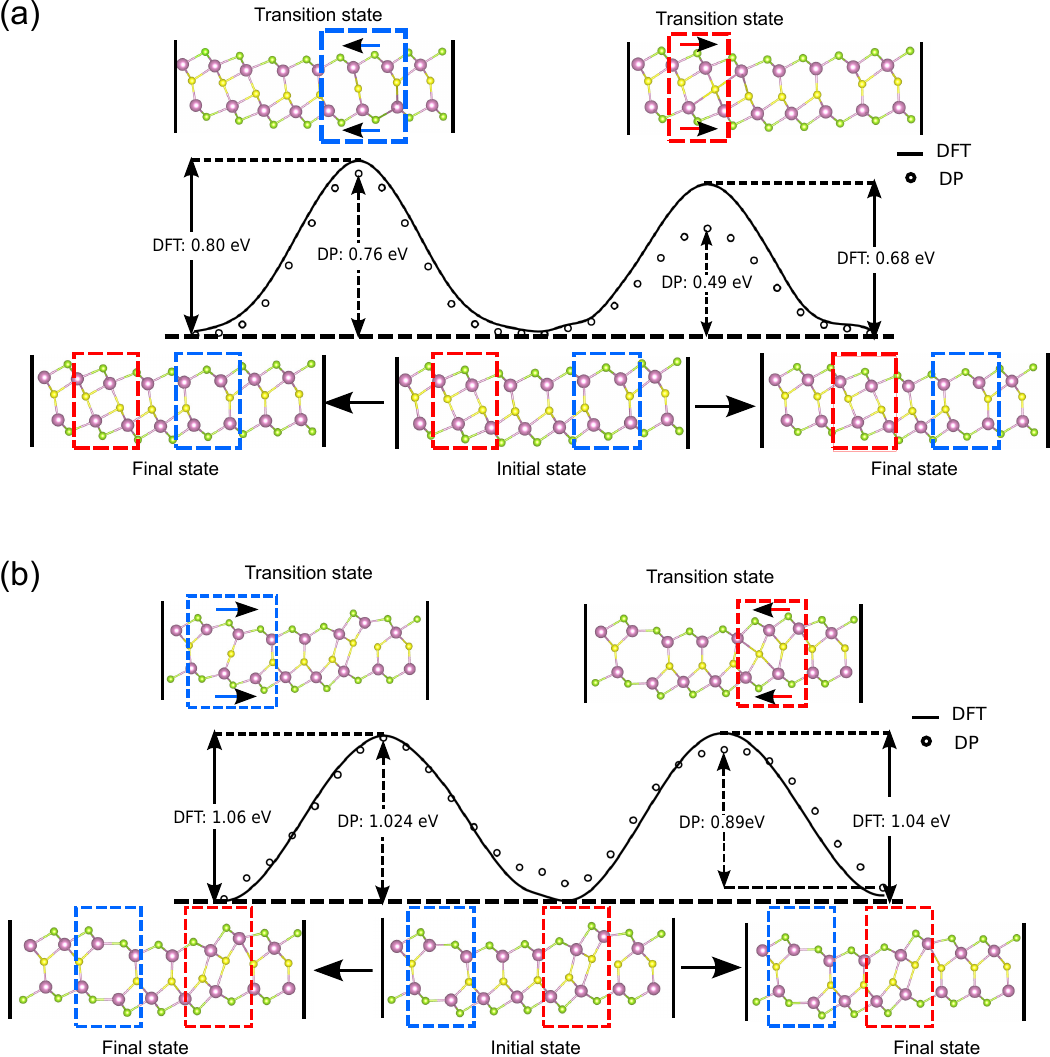} 
\caption{DFT and DP energy profiles for 180$^\circ$ domain wall motions in monolayer $\alpha$-In$_2$Se$_3$. Because of the in-plane lattice asymmetry, there are four different types of 180$^\circ$ domain walls. The initial pathways were taken from ref.~\citenum{Ding17p14956} and were re-optimized using a higher planewave cutoff energy and more intermediate images.}
\label{fir:DW}
\end{figure}

For bulk ferroelectrics, it is well established that the electric field-driven domain wall motion plays a critical role in ferroelectric switching. It is thus essential for the DP model to reproduce the energy barrier of lateral domain wall motion. 
Because of the in-plane lattice asymmetry, there are four different types of 180$^\circ$ domain walls in monolayer $\alpha$-In$_2$Se$_3$. The initial pathways of 180$^\circ$  domain wall motions were taken from ref.\citenum{Ding17p14956} and were re-optimized using a higher planewave cutoff energy and more intermediate images. As shown in Fig.~\ref{fir:DW}, the DP energy profiles again agree with DFT results for all four pathways.

\subsection{Phase transitions}
\begin{figure}  
\centering
\includegraphics[width=0.8\textwidth]{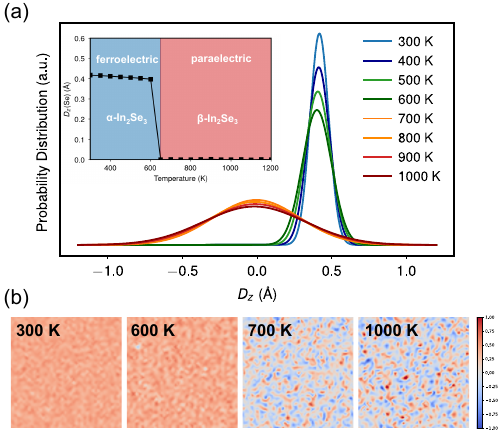} 
\caption{Temperature-driven ferroelectric-paraelectric phase transition in monolayer $\alpha$-In$_2$Se$_3$. (a) Temperature-dependent probability distributions of atomic displacements ($D_z$) of central Se sublayer as a function of temperature. The inset shows the evolution of $D_z$ with increasing temperature, confirming a $\alpha \rightarrow \beta$ transition at 650~K. (b) $D_z$ maps for instantaneous configurations sampled at different temperatures.}
\label{fir:dis-1D}
\end{figure}

Our main purpose is to obtain an accurate force field for MD simulations of monolayer {In$_2$Se$_3$}. We simulate the temperature-driven phase transition starting with the ferroelectric $\alpha$ phase using the DP model and a 8000-atom supercell. The out-of-plane displacement ($D_z$) of the central Se atoms with respect to the geometric center of the monolayer is used to measure the local symmetry breaking. Figure \ref{fir:dis-1D}(a) shows the evolution of the probability distribution of $D_z$ with increasing temperature. At each temperature, the distribution of $D_z$ can be fitted with a single Gaussian. The peak position shifts towards lower values and sharply decreases to zero when the temperature changes from 600~K to 650~K (see Fig.~\ref{fir:dis-1D}(a) inset), which reveals a ferroelectric-to-paraelectric ($\alpha \rightarrow \beta $) phase transition (see the trajectory movies of $D_z$ maps in a public repository)~\cite{GroupGithub}. The value of $T_C$ from DPMD simulations is $\approx650$~K, agreeing with previous first-principles MD studies ($\approx650$~K)~\cite{Liu19p025001} and experimental value of 700~K in 3-nm-thick $\alpha$-In$_2$Se$_3$ crystals~\cite{Xiao18p227601}. Our large-scale MD simulations suggest that the $\alpha \rightarrow \beta $ transition is mostly of the displacive character~\cite{Qi16p134308,Wexler10p174109}.  We also plot the 2D maps of $D_z$ for instantaneous configurations sampled at different temperatures in Fig.~\ref{fir:dis-1D}(b). Consistent with the probability distribution of $D_z$, the values of $D_z$ also show considerable spacial variations when $T>T_C$.

The Steinhardt's bond orientational order parameters ($Q_l$)~\cite{Steinhardt83p784,Mickel13p044501} are most commonly used to distinguish different crystalline phases and clusters of particles in 3D. The $Q_l$ at site $i$ is calculated according to 
\begin{equation}  
      Q_l=\sqrt{(\frac{4\pi}{2l+1})\sum_{m=-l}^{m=l}{\bar{Y}_{lm}}{\bar{Y}_{lm}}^*}
\end{equation}

\begin{equation}  
      {\bar{Y}_{lm}}= \frac{1}{N_b(i)}\sum_{j\in {N_b(i)}}{Y_{lm}}(\theta(r_{ij}),\phi(r_{ij}))
\end{equation}
where $l$ is the order of symmetry, ${\bar{Y}_{lm}}$ is the average of the spherical harmonics ${Y_{lm}}(\theta(r_{ij}), \phi(r_{ij}))$ with $\theta(r_{ij})$ and $\phi(r_{ij})$ being the azimuthal and polar angles of the bond pointing from particle $i$ to particle $j$, respectively, and ${N_b(i)}$ is the number of neighbors of atom $i$. It is easy to see that $Q_l$ is rotational invariant and non-negative.
Table II reports the layer-resolved $Q_4$ and $Q_6$ values of Se atoms in $\alpha$ and $\beta$ phases using DFT optimized structures. It shows that Se atoms have distinct $Q_4$ and $Q_6$ values in these two phases. In particular, $Q_4$ of Se atoms of the middle layer in the $\beta$ phase (0.082) is almost twice the value than that in the $\alpha$ phase (0.049). Because of the out-of-plane polarization, there are three symmetrically unique Se atoms in the $\alpha$ phase, corresponding to different $Q_4$ and $Q_6$ values.

\begin{figure}  
\centering
\includegraphics{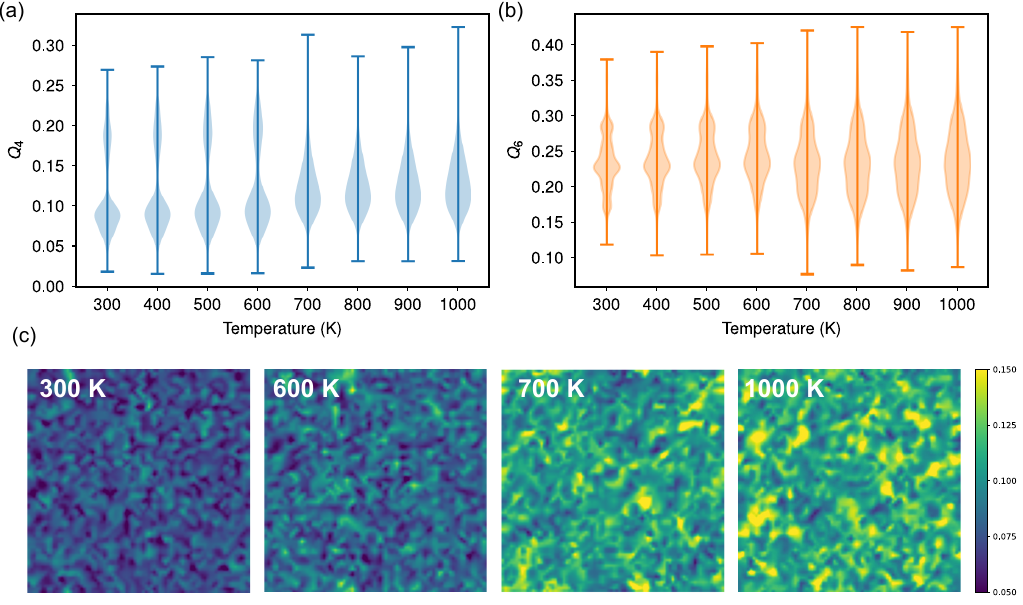} 
\caption{Violin plot of (a) $Q_4$ and (b) $Q_6$ of Se atoms in monolayer $\alpha$-In$_2$Se$_3$ as a function of temperature. (c) 2D map of $Q_4$ for Se atoms in the middle sublayer for instantaneous configurations sampled at different temperatures.}
\label{fir:violin}
\end{figure}

We calculate $Q_4$ and $Q_6$ of all the Se atoms for configurations sampled at different temperatures and present their violin plots in Fig.~\ref{fir:violin}(a). The density curves of $Q_4$ at $T<T_C$ exhibit two peaks, with the higher value peak ($\Lambda_h$) corresponding to surface Se atoms and the lower value peak ($\Lambda_l$) coming from central Se atoms, and the value of $\Lambda_l$ increases with temperature. At $T>T_C$, the density curves become single peaked. Overall, $Q_4$ serves as a good indicator of local atomic environment and $\alpha \rightarrow \beta$ phase transition. The density curves of $Q_6$ change from distributions possessing multiple peaks below $T_C$ to unimodal distributions above $T_C$. Finally,  we present 2D maps of $Q_4$ for Se atoms in the middle sublayer for instantaneous configurations sampled at selected temperatures (Fig.~\ref{fir:violin}(b)). Similar to the 2D $D_z$ maps shown in Fig.~\ref{fir:dis-1D}(b), $Q_4$ maps also reveal substantial structural heterogeneity in space particularly at temperatures above $T_C$.

\section{Conclusions}
Using the concurrent learning scheme, we developed a DP model for monolayer In$_2$Se$_3$. The model potential exhibits accuracy comparable with first-principles methods, as confirmed by the excellent agreement between DFT and DP results for various thermodynamic properties of In$_2$Se$_3$ polymorphs and kinetic pathways of polarization reveal and 180$^\circ$ domain wall motions in $\alpha$-In$_2$Se$_3$. The constant-temperature constant-pressure MD simulations predict a temperature-driven $\alpha \rightarrow \beta$ phase transition at the single-layer limit. Both local atomic displacement and Steinhardt's bond orientational order parameter $Q_4$ are used to characterize the phase transition, revealing a displacive feature.  The DP model developed in this work could be a starting point for research of dynamical processes in $\alpha$-In$_2$Se$_3$-based nanoelectronics. Our work demonstrates the robustness of DP for 2D materials. Following a similar protocol we suggested in our previous work~\cite{Wu21p024108}, we make our final training database and hyperparameters available through a public repository~\cite{DPLibrary}, which can be used for training a new DP model or any other ML-based force field based on desired exchange-correlation functionals such as SCAN that has demonstrated accuracy for solid-state materials~\cite{Sun15p036402,Isaacs18p63801,Wexler20p054101}.
We expect the current DP model can be systematically improved and extended, enabling large-scale simulations of interlayer sliding ferroelectricity and Moir\'e ferroelectricity of vdW multilayers. Specifically, we can
add new configurations of bilayer In$_2$Se$_3$ of different phases to the current training database and set up exploration runs of various bilayer systems at different temperatures to generate new configurations in the DP-GEN scheme. After a converged training with a database containing diverse configurations of layered systems, we believe the DP model, thanks to the ability of the deep neural network to faithfully represent complex energy landscape, can accurately describe the subtle changes in energy and atomic forces during the slide and rotation of two monolayers, essential for the modeling of interlayer sliding ferroelectricity and Moir\'e ferroelectricity.

 

\section{ACKNOWLEDGMENTS}
J.W.,L.B.,J.H., L.M., and S.L.acknowledge the support from Westlake Education Foundation, Westlake Multidisciplinary Research Initiative Center, and the National Natural Science Foundation of China (52002335). The computational resources were provided by Westlake HPC Center. The authors thank Dr. Wenjun Ding for sharing the trajectories of polarization reversal and domain wall motions in moonolayer $\alpha$-In$_2$Se$_3$. The authors also thank Zhe Wang and Dr. Wenguang Zhu for providing the structure file of $\beta_3'$-In$_2$Se$_3$.

\bibliography{SL}

\newpage
\begin{table}[ht]
\centering
\caption{DFT and DP energy ($E$) in eV per 5-atom unit cell for monolayer In$_2$Se$_3$ polymorphs in reference to monolayer $\alpha$-In$_2$Se$_3$. Lattice constants $a$ and $b$ in \r{A} are optimized with DFT. The $\alpha$, $\beta$, $\beta'$ phases have hexagonal unit cells while $\beta_1'$, $\beta_2'$, $\beta_3'$ have orthorhombic unit cells.}
\label{Telas}
\scalebox{1.0}{
\centering
\begin{ruledtabular}
\begin{tabular}{lcccccc} 
&$\alpha$ & $\beta_3'$&  $\beta_1'$  & $\beta'$ & $\beta_2'$  &$\beta$ \\
\hline
$a$& 4.063 &8.093 & 4.093  &3.963   &3.962 & 3.982 \\
$b$& 4.063  &7.045 & 6.816 & 4.040 & 6.816& 3.982 \\
$E_{\rm DFT}$ & 0 & 0.0308 & 0.0483 & 0.0523  &  0.0527 & 0.1485\\
$E_{\rm DP}$ & 0 & 0.0157 & 0.0461 & 0.0521 & 0.0525 & 0.1703 \\
\end{tabular}
\end{ruledtabular}
}
\end{table}

\begin{table}[ht]
\centering
\caption{Steinhardt’s bond orientational order parameters ($Q_l$) for Se atoms in monolayer In$_2$Se$_3$ of $\alpha$ and $\beta$ phases. The $\alpha$ phase has polarization pointing downward as shown in Fig.~1. }
\label{Telas}
\scalebox{1.0}{
\centering
\begin{ruledtabular}
\begin{tabular}{lcccccc} 
&$Q_4^{\rm top}$ & $Q_4^{\rm mid}$& $Q_4^{\rm bottom}$ &  $Q_6^{\rm top}$  & $Q_6^{\rm mid}$ & $Q_4^{\rm bottom}$ \\
\hline
$\alpha$& 0.073 &0.049& 0.141  &0.163&0.175 & 0.245\\
$\beta$&0.116 &0.082 & 0.082 &0.101 & 0.166& 0.166\\

\end{tabular}
\end{ruledtabular}
}
\end{table}

\newpage
\newpage

\newpage

\newpage
\newpage

\newpage
\newpage

\newpage
\newpage

\clearpage
\newpage
\newpage

\newpage

\clearpage
\newpage

\clearpage
\newpage

\newpage
\clearpage

\end{document}